\documentclass[aps,pra,reprint,showpacs]{revtex4-1}

\usepackage{physics}
\usepackage{graphicx}
\usepackage{amsfonts}
\usepackage{amssymb}
\usepackage[mathscr]{eucal}

\begin{document}

\title{Steady state thermodynamics of two qubits strongly coupled to bosonic environments}
\author{Ketan Goyal}
\altaffiliation{Current Address: Avigo Solutions, LLC 1500 District Avenue, Burlington, MA 01803, USA}
\author{Ryoichi Kawai}
\affiliation{Department of Physics, University of Alabama at Birmingham, Birmingham AL 35294, USA}

\begin{abstract}
When a quantum system is placed in thermal environments, we often assume that the system relaxes to the Gibbs state in which decoherence takes place in the system energy eigenbasis.   However, when the coupling between the system and the environments is strong, the stationary state is not necessarily the Gibbs state due to environment-induced decoherence which can be interpreted as continuous measurement by the environments.  Based on the einselection proposed by Zurek, we postulate that the Gibbs state is projected onto the pointer basis due to the continuous measurement. We justify the proposition by exact numerical simulation of a pair of coupled qubits interacting with boson gases. Furthermore, we demonstrate that heat conduction in non-equilibrium steady states can be suppressed in the strong coupling limit also by the environment-induced decoherence. 

\end{abstract}

\date{\today}


\maketitle

\section{Introduction}

The laws of thermodynamics and the principles of statistical mechanics tell us that every system eventually reaches a stationary state  known as the Gibbs state, which is the hallmark of thermal equilibrium.  The density operator of the Gibbs state is notably a function of only the system Hamiltonian and is thus diagonal in the energy eigenbasis. The coherence between energy eigenstates is completely destroyed.  Therefore, thermalization to the Gibbs state must involve decoherence between energy eigenstates, presumably induced by the environments surrounding the system. 

Such a decoherence process toward the Gibbs state has been investigated under the weak coupling limit\cite{vanHove1957}. In fact, quantum master equations based on the Born-Markovian approximation are known to converge to the Gibbs state.\cite{Breuer2002}.  However, it has been shown that the non-Markovian dynamics does not necessarily reach the Gibbs state.\cite{Mori2008,Genway2012,Lee2012,Cai2014,Xiong2015,Vega2017}  For a system strongly coupled to the environments, its equilibrium state cannot be expressed with the system Hamiltonian alone, and an effective Hamiltonian based on the potential of mean force has been developed.\cite{Gelin2009,Campisi2010,Hilt2011,Esposito2015,Seifert2016,Jarzynski2017,Miller2017,Strasberg2019}  The resulting stationary state is no longer diagonal in the system energy eigenstates.

  \begin{figure}[b]
     \includegraphics[width=3.3in]{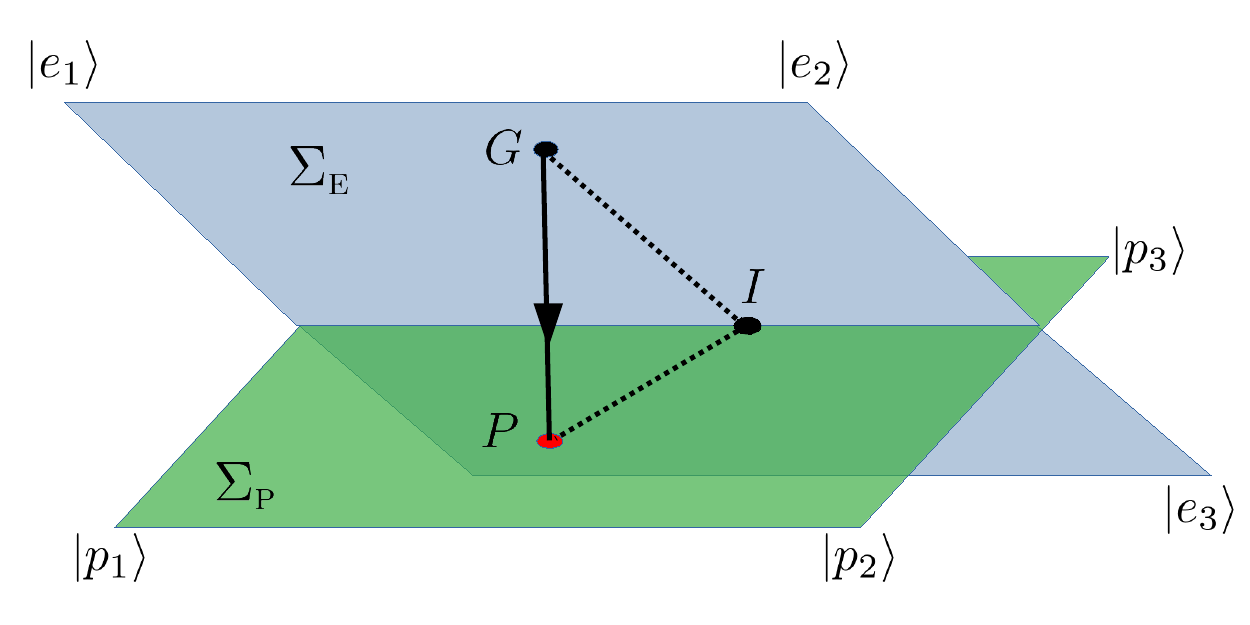}
     \caption{Schematic representation of Proposition (\ref{eq:proposition}). The Gibbs state on the convex hull $\Sigma_\textsc{e}$ is projected onto another convex hull $\Sigma_\textsc{p}$.  As the coupling strength increases, the steady state deviates from the Gibbs state ($G$) along the projection line toward the pointer limit ($P$). The maximal entropy state ($I$) is located on the intersection of the two convex hulls.  Noting that $P$ is closer to $I$ than $G$, the entropy increases as the steady state moves toward the pointer limit.}
     \label{fig:projection}
  \end{figure} 

Environment-induced decoherence has been intensively investigated  in the context of quantum measurement theory and quantum computing.\cite{Schlosshauer2007}  In those theories, the environment does not necessarily induce decoherence in the energy eigenbasis.  Zurek\cite{Zurek2003,Eisert2004} showed that the decoherence takes place among so-called ``pointer states'' determined by the coupling Hamiltonian between a system and environments. In general, the system density operator becomes diagonal in the pointer basis under the strong coupling limit. This \emph{einselection}\cite{Zurek2003} can be considered as a consequence of continuous measurement of the system by the environment. A similar argument can be used for the thermalization processes, and there have been investigation of thermalization under continuous measurement.\cite{Jack2000,Ashida2018}   We investigate thermalization and heat conduction in the strong coupling regime based decoherence in the pointer basis.

\section{Thermalization in the Poiter Basis}
Consider a system in the Gibbs state $\rho_\textsc{s}^\textsc{g}=e^{-\beta H_\textsc{s}}/Z_\textsc{s}$ under the weak coupling, where $H_\textsc{s}$  $\beta$, and $Z_\textsc{s}$  are system Hamiltonian, inverse temperature and a partition function.  When the coupling energy becomes significantly larger than the system energy, the Gibbs state is continuously measured by the environments and thus projected to the pointer basis. Our main proposition is that under the strong coupling limit a system tends to relax to a stationary state  given by
\begin{equation}\label{eq:proposition}
\rho_\textsc{s} \xrightarrow[]{t \rightarrow \infty} \frac{1}{Z_\textsc{s}} \sum_i  \dyad{p_i} \rho_\textsc{s}^\textsc{g} \dyad{p_i}
\end{equation}
where  $\ket{p_i}$ is the $i$-th pointer state which we define below.  

Figure \ref{fig:projection} illustrates this proposition.  Consider the convex hull $\Sigma_\textsc{e} = \left\{ \rho=\sum_i Q_i \dyad{e_i} ;\, Q_i \ge 0 \wedge \sum_i Q_i = 1 \right\}$ in the Liouville space.  The corners of the hull represent the pure states.  Any density operator that is diagonal in the energy eigenbasis $\ket{e_i}$ is in $\Sigma_\textsc{u}$, including the Gibbs state ($G$ in Fig \ref{fig:projection}).  Similarly, the convex hull  $\Sigma_\textsc{p} = \left\{ \rho=\sum_i P_i \dyad{p_i};\, P_i \ge 0 \wedge \sum_i P_i = 1 \right\}$ contains all possible density operators that are diagonal in the pointer basis $\ket{p_i}$.  The density operators in the intersection of the two convex hulls are diagonal in both basis sets.  A special point $I$ in the figure corresponds to $\rho = \frac{1}{d_\textsc{s}} I_\textsc{s}$ where $I_\textsc{s}$ is an identity operator and $d_\textsc{s}$ is the dimension of the system Hilbert space.  The entropy of the system reaches its maximum  value $\ln d_\textsc{s}$ at $I$.  As the coupling gets stronger, the steady state deviates from the Gibbs state ($G$) toward the pointer limit ($P$) along the projection line ($\overline{GP}$).   The projection line is ``perpendicular'' to $\Sigma_\textsc{p}$, meaning that the diagonal elements in the pointer basis are invariant along the projection line.

\section{Model and Numerical Simulation}
We justify the proposition by numerically investigating the exact dynamics of a simple spin-boson model.
Following the standard open quantum system approach\cite{Breuer2002}, we consider an isolated system
consisting of a small subsystem $\mathcal{H}_\textsc{s}$ and environments  
$\mathcal{H}_\textsc{b}$.  
The unitary evolution of the total system follows the Liouville--von Neumann equation
\begin{equation}\label{eq:eom_total}
i \pdv{t}\rho_\textsc{sb} = 
\comm{H_\textsc{s}+ H_{\textsc{b}}+V_{\textsc{sb}}}{\rho_\textsc{sb}}.
\end{equation}
where $H_\textsc{b}$ is the Hamiltonian of environment. For simplicity, we assume that the coupling Hamiltonian takes a bilinear form  \begin{equation}
   V_\textsc{sb} = \sum_\ell X_\ell \otimes Y_\ell
\end{equation}
where $X_\ell$ and $Y_\ell$ are operators in $\mathcal{H}_\textsc{s}$ and  $\mathcal{H}_\textsc{b}$, respectively.  Furthermore, we assume that $\comm{X_k}{X_\ell}=0$ so that all $X_\ell$ share the same eigenkets $\ket{p_j}$ which we shall call pointer states. If there are degenerate subspaces, we choose a particular basis in the subspace such that the steady state becomes diagonal in the pointer basis.

The state of the system is represented by reduced density $\rho_\textsc{s} = \Tr_\textsc{b} \rho_\textsc{sb}$ which obeys the equation of motion
\begin{equation}\label{eq:eom_system}
    i \dv{t}\rho_\textsc{s} = \comm{H_\textsc{s}}{\rho_\textsc{s}} + \sum_\ell \comm{X_\ell}{\eta_\ell}
\end{equation}
where we introduced a new operator,
\begin{equation}\label{eq:eta_def}
\eta_\ell \equiv  \Tr_\textsc{b} \left \{ \rho_\textsc{sb} Y_\ell  
\right \} \quad \in \mathcal{H}_\textsc{s} .
\end{equation}
Note that the time evolution of the system needs only limited information on the state of the environments through $\eta_\ell$.

In order to demonstrate the proposition, we consider a simple model consisting of a 
pair of identical qubits S$_{1}$ and S$_{2}$ whose Hamiltonian is given by 
\begin{equation}
H_\textsc{s} = \frac{\omega_0}{2} \sigma^{z}_{1}	+ 
\frac{\omega_0}{2} \sigma^{z}_{2} + \lambda_\textsc{s} \left(\sigma^{+}_{1} 
\sigma^{-}_{2} + \sigma^{-}_{1} \sigma^{+}_{2} \right )
\end{equation} 
where $\sigma_{\ell}^{z,\pm},(\ell=1,2)$ are usual Pauli matrices for the $\ell$-th qubit, and $\omega_0$ and $\lambda_\textsc{s}$ the qubit excitation energy and the internal coupling strength, respectively.  We write the energy eigenstates as $\ket{e_j},\,(j=1,\cdots,4)$ with eigenvalue $e_j$ starting from the ground state.

Each qubit S$_{\ell}$ is coupled to its own environment  B$_{\ell}$.\footnote{If two qubits share the same environment, decoherence-free subspaces could be formed, which is protected from decoherence due to symmetry.  We avoid the decoherence free subspace by using two independent environments.}
The environments are assumed to be ideal Bose gases whose Hamiltonians are given by $H_{\textsc{b}_\ell} 
= \sum_k 
\omega_{\ell}(k)\, a^\dagger_{\ell}(k) a_{\ell}(k)$, where $a^\dagger_\ell(k)$ 
and 
$a_\ell(k)$ are creation and annihilation operators for the $k$-th mode in $B_\ell$. The interaction Hamiltonian between S$_{\ell}$ and B$_{\ell}$ assumes a simple bilinear form $X_{\ell} \otimes Y_{\ell}$
where $X_{\ell}=\sigma^{x}_{\ell}$ and $Y_{\ell} = 
\sum_k \epsilon_{\ell}(k) \left[a^\dagger_{\ell}(k) + a_{\ell}(k) \right]$.  The coupling strength between the system and the $k$-th mode in $B_\ell$ is denoted as $\epsilon_{\ell}(k)$.   

\begin{figure}
   \centering
   \includegraphics[width=3.3in]{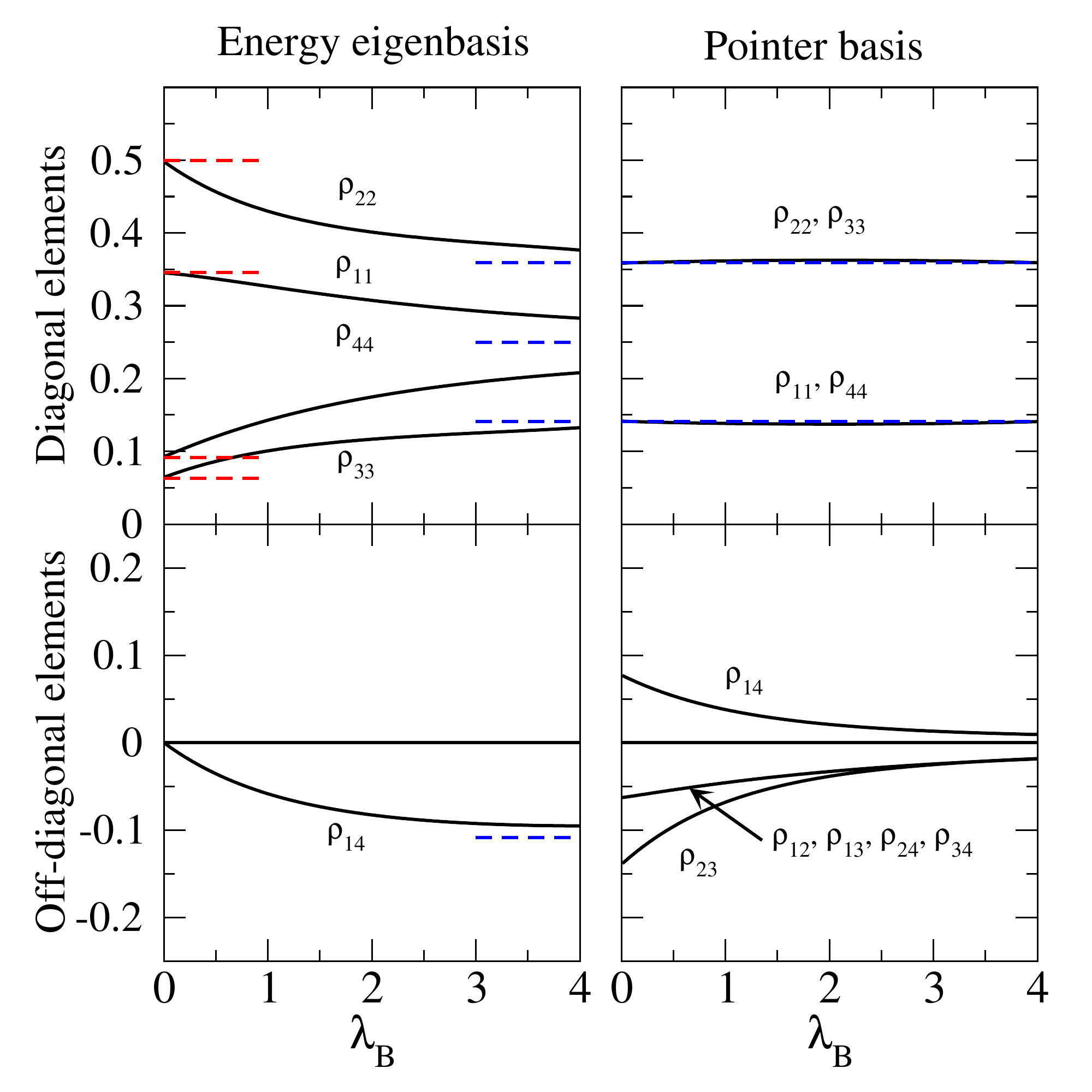}
   \caption{Stationary state density matrix, diagonal (top) and off-diagonal elements (bottom), are plotted as a function of coupling strength $\lambda_\textsc{b}$.  In the left panel the matrix is evaluated in the eigenbasis of the system Hamiltonian $H_\textsc{s}$ and in the right panel the pointer basis is used.  The parameter values $\omega_0=1$, $\lambda_\textsc{s}=1.55$, $T=1.5$, $\gamma_\textsc{b}=0.15$ are used. The Gibbs density matrix in the energy eigenbasis is shown as red dashed lines, and the strong coupling limit (pointer state limit) predicted by the present proposition is shown as blue dashed lines.}
   \label{fig:equilibrium}
\end{figure}

The pointer states in this model are the simultaneous eigenkets of $X_1$ and $X_2$ and denoted as $\ket{p_1}=\ket{0\,0}$, $\ket{p_2}=\ket{0\,1}$, $\ket{p_3}=\ket{1\,0}$, and $\ket{p_4}=\ket{1\,1}$, where $\ket{0}$ and $\ket{1}$ are the eigenkets of $\sigma^{x}$.

When the coupling is weak, the stationary state is the Gibbs state
\begin{equation}\label{eq:gibbs}
   \rho_\textsc{s}  \xrightarrow[]{t \rightarrow \infty}  \sum_j \rho^\text{e}_{jj} \dyad{e_j}
\end{equation}
where $\rho^\text{e}_{jj} = e^{-\beta e_j} /\sum_i e^{-\beta e_i}$.  Under the strong coupling limit,
Proposition (\ref{eq:proposition}) claims that the stationary state density is given by
\begin{equation}\label{eq:pointer_limit}
    \rho_\textsc{s} \xrightarrow[]{t \rightarrow \infty} \sum_j \rho_{jj}^\text{p} \dyad{p_j}
\end{equation}
where $\rho^p_{jj} = \mel{p_j}{\rho_\textsc{s}^\textsc{g}}{p_j}/Z_\textsc{s}$ can be explicitly expressed as 
\begin{subequations}\label{eq:pointer_limit_matrix}
\begin{eqnarray}
    \rho_{11}^\text{p}=\rho_{44}^\text{p} &=& \frac{1}{4}\left ( 1- 
    \frac{\sinh\beta\lambda_\textsc{s}}{\cosh\beta \omega_0 + \cosh\beta 
    \lambda_\textsc{s}} \right )\\
    \rho_{22}^\text{p}=\rho_{33}^\text{p} &=& \frac{1}{4}\left ( 1+ 
    \frac{\sinh\beta\lambda_\textsc{s}}{\cosh\beta \omega_0 + \cosh\beta 
    \lambda_\textsc{s}} \right )
   \end{eqnarray}    
\end{subequations}

Now we show the transition from the Gibbs limit (\ref{eq:gibbs}) to the pointer limit (\ref{eq:pointer_limit}) by numerically solving Eq. (\ref{eq:eom_system}). 
Assuming that the total system is initially in a product state $\rho(t_0) = \rho_\textsc{s} (t_0) \otimes \rho_\textsc{b} (t_0)$ with the environment in a thermal state $\rho_\textsc{b}(t_0) = \prod_\ell \exp(-\beta_\ell H_{\textsc{b}_\ell})/Z_{\textsc{b}_\ell}$, we obtain a formally exact expression of the system density in the interaction picture\cite{Ishizaki2009}
   \begin{equation}\label{eq:rho_s_exact}
    \rho_\textsc{s}(t) = \overleftarrow{\mathscr{T}} \prod_{\ell}  
    e^{-\int_{t_0}^{t} 
    \int_{t_0}^{t_1} \dd{t_1} \dd{t_2} \mathscr{K}_{\ell}(t_1,t_2)} 
    \rho_\textsc{s}(t_0)
\end{equation}
where the super operator $\mathscr{K}_{j}$ is defined by
\begin{eqnarray}\label{eq:memory_kernel}
    \mathscr{K}_{\ell}(t_1,t_2) &=& \mathscr{S}^{-}_{\ell}(t_1)\,  
    K^\text{(n)}_{\ell}(t_1-t_2)\,  \mathscr{S}^{-}_{\ell}(t_2) \nonumber\\
    & &+ i \mathscr{S}^{-}_{\ell}(t_1)\, K^\text{(d)}_{\ell}(t_1-t_2), 
    \mathscr{S}^{+}_{\ell}(t_2)
\end{eqnarray}
with anti(+) and regular(-) commutators $\mathscr{S}^\pm_\ell=\comm{X_\ell}{\cdot}_\pm$.   The 
dissipation kernel $K^\text{(d)}_{\ell}(t)$ and noise kernel $K^\text{(n)}_{\ell}(t)$ 
are respectively the real and imaginary part of the correlation function 
$C_\ell(t)=\expval{Y_{\textsc{b}_\ell}(t) Y_{\textsc{b}_\ell}(t_0)}_0$ where the expectation value is taken 
with the initial environment state $\rho_{\textsc{b}_\ell}(t_0)$. The time 
ordering operator $\overleftarrow{\mathscr{T}}$ in Eq. (\ref{eq:rho_s_exact}) chronologically orders the 
super operators $\mathscr{S}^{\pm}_{\ell}(t)$.

Kato and Tanimura\cite{Kato2015,*Kato2016} showed that Eq (\ref{eq:rho_s_exact}) can be numerically evaluated if  the spectral density of environments is of the Drude--Lorentz type:
\begin{equation}
g_{\ell}(\omega)= 
\frac{2 \lambda_{\ell} \gamma_{\ell} \omega}{\omega^2+\gamma_{\ell}^2} 
\end{equation}
where $\gamma_{\ell}$ and $\lambda_{\ell}$ are the response rate of environment and 
the overall coupling strength between qubit $S_{\ell}$ and environment $B_{\ell}$, respectively.
Then, the environmental correlation can be expressed with reasonable accuracy as\cite{Xu2009}
\begin{eqnarray}
C_{\ell}(t) &\approx&  \lambda_{\ell} \left [ c_{\ell}\, e^{-\gamma_{\ell}} + 2 \Delta_{\ell}\, \delta(t) \right ]
\end{eqnarray}
where $c_{\ell} = 2/\beta_{\ell} -\gamma_{\ell} \Delta_{\ell} - i \gamma_{\ell}$ and $\Delta_{\ell} = \gamma_{\ell} \beta_{\ell} / 6$. 

\begin{figure}
    \centering
    \includegraphics[width=2.5in]{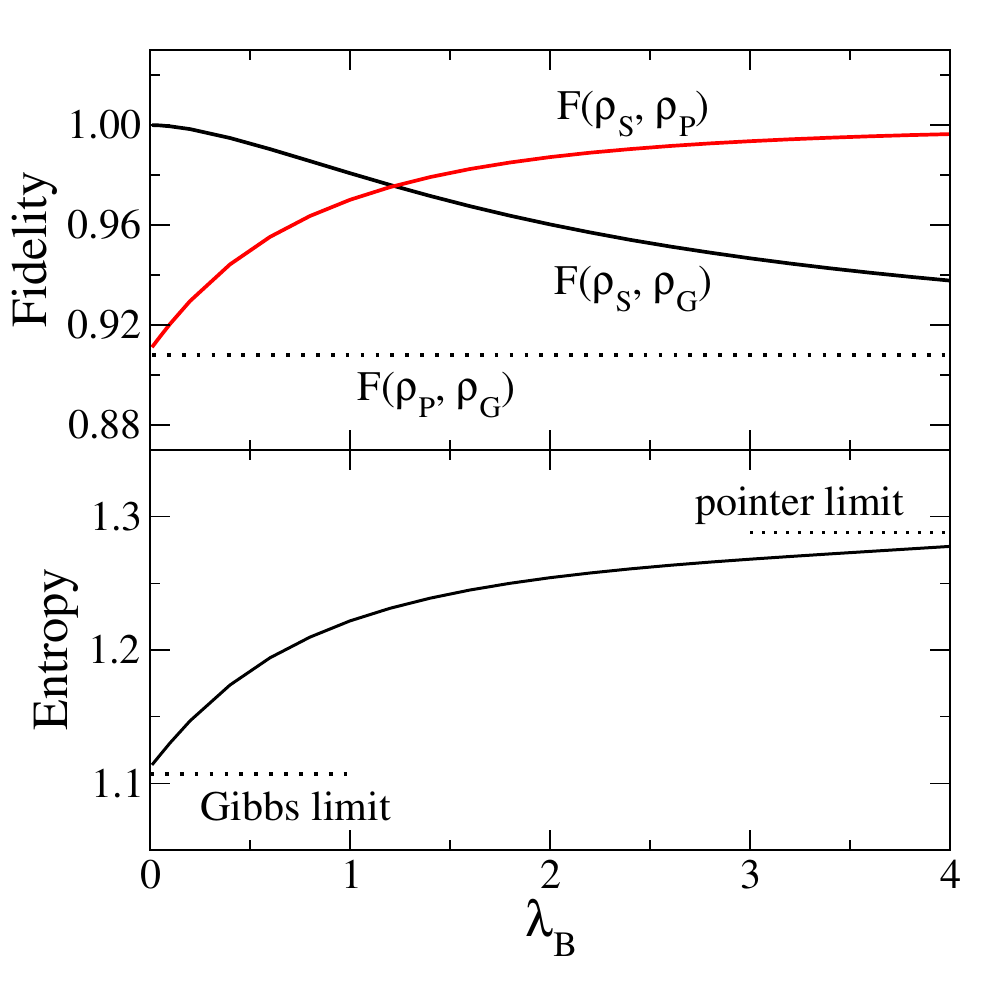}
    \caption{The fidelity (upper panel) between the steady state $\rho_\textsc{s}$, the Gibbs state $\rho_\textsc{g}$, and the pointer limit $\rho_\textsc{p}$ shows that the steady state deviates from the Gibbs state and approaches the pointer limit. The entropy of the steady (lower panel) state also deviates from the Gibbs limit and approaches the pointer limit. See Fig. \ref{fig:equilibrium} for the parameter values.} \label{fig:entropy-fidelity}
\end{figure}

Following Kato and Tanimura\cite{Kato2015,Kato2016},  we introduce a set of auxiliary operators
\begin{multline}
\zeta_{n_1,n_2}(t) =  \overleftarrow{\mathscr{T}}  \prod_{\ell} 
\left \{ \left[ -i\int_{t_0}^{t} \dd{s} e^{-\gamma_{\ell} (t-s)} \mathscr{G}_{\ell} (s) \right]^{n_\ell} \right . \\
\left . \times  e^{- \lambda_{\ell} \int_{t_0}^{t} \int_{t_0}^{t_1} \dd{t_1} \dd{t_2} 
    \mathscr{S}^{-}_{\ell}(t_1) e^{-\gamma_\ell(t_1-t_2)} \mathscr{G}_{\ell}(t_2) } \right . \\
\left . \times  e^{- \lambda_{\ell} \Delta_{\ell} \int_{t_0}^{t} \dd{t_1} \mathscr{S}^{-}_{\ell}(t_1) \mathscr{S}^{-}_{\ell}(t_1) } \right \} \rho_\textsc{s}(t_0) 
\end{multline}
where 
\begin{equation}
\mathscr{G}_{\ell}(t) =\left ( 2/\beta_{\ell} - \gamma_{\ell} \Delta_{\ell} \right ) \mathscr{S}^{-}_{\ell}(t) - i \gamma_{\ell} \mathscr{S}^{+}_{j}(t).
\end{equation}
Index $n_\ell$ associated with environment $B_\ell$ runs from 0 through infinity. Only the first three lowest order auxiliary operators are needed for   $\rho_\textsc{s}(t) = \zeta_{0,0}(t)$, $\eta_1 = \lambda_{1} \left [\zeta_{1,0}(t) - i \Delta_{1} \mathscr{S}^{-}_{1}(t) \zeta_{0,0}(t) \right ]$ and $\eta_2 = \lambda_{2} \left [\zeta_{0,1}(t) - i \Delta_{2} \mathscr{S}^{-}_{2}(t) \zeta_{0,0}(t) \right ]$.  However, 
the dynamics of auxiliary operators is determined by an infinite set of coupled ODEs or so-called  hierarchical equations of motion\cite{Kato2015,Kato2016}
\begin{multline}\label{eq:heom-general}
\dv{t} \zeta_{n_{1},n_{2}} (t) =  -(\gamma_{1} n_{1} + \gamma_{2} n_{2})\, \zeta_{n_{1},n_{2}}(t) \\
-\left [\lambda_{1}\Delta_{1} \mathscr{S}^{-}_{1}(t)\mathscr{S}^{-}_{1}(t)+\lambda_{2}\Delta_{2} \mathscr{S}^{-}_{2}(t)\mathscr{S}^{-}_{2}(t) \right ]\,  \zeta_{n_1,n_2}(t) \\
-i\lambda_{1} \left [\mathscr{S}^{-}_{1}\zeta_{n_{1}+1,n_{2}}(t) - n_{1} \mathscr{G}_{1}(t) \zeta_{n_{1}-1,n_{2}}(t)\right ] \\
-i\lambda_{2} \left [\mathscr{S}^{-}_{2}\zeta_{n_{1},n_{2}+1}(t) - n_{2} \mathscr{G}_{2}(t) \zeta_{n_{1},n_{2}-1}(t)\right]
\end{multline}
with the initial condition $\zeta_{n_1, n_2}(t_0) = 0$ except for $\zeta_{0, 0}(t_0) = \rho_\textsc{s}(t_0)$.   The infinite hierarchy is truncated at depth $d=50$ such that higher depth auxiliary operators do not significantly  contribute to the first two depths.   

\section{Results and Discussion}

First, we investigate the equilibrium situation where the initial states of the two environments are identical ($\lambda_{1}=\lambda_{2}\equiv \lambda_\textsc{b}$, $T_{1}=T_{2}\equiv T$, $\gamma_1=\gamma_2\equiv \gamma$). We tried more than ten different initial densities, and all converged to the same stationary state.
In Fig. \ref{fig:equilibrium}, the matrix elements of the stationary state density are plotted as a function of the coupling strength $\lambda_\textsc{b}$ using the energy eigenbasis and the pointer basis.
The density matrix in the energy eigenbasis shows that the Gibbs state is realized only at the weak coupling limit. The diagonal elements deviate from the Gibbs state as the coupling increases. The off-diagonal elements indicate that the superposition of eigenstates $\ket{e_1}$ and $\ket{e_4}$ grows rapidly and thus decoherence does not fully take place in the energy eigenbasis.  Both the diagonal and off-diagonal elements approach the pointer limit predicted by Eq. (\ref{eq:pointer_limit}).

When the matrix elements of the same density operator are evaluated in the pointer basis, all of the off-diagonal elements tend to vanish as the coupling strength increases, suggesting that full decoherence takes place in the pointer basis.  The diagonal elements are remarkably insensitive to the coupling strength and in good agreement with Eq. (\ref{eq:pointer_limit_matrix}) regardless of the coupling strength.  The invariance of the diagonal elements confirms that the projection is perpendicular to the convex hull $\Sigma_\textsc{p}$. (See Fig. \ref{fig:projection}.) In  Fig. \ref{fig:entropy-fidelity} the deviation of the steady state from the Gibbs state and its approach to the pointer limit are measured by fidelity $F(\rho,\rho') = \left ( \tr \{ \sqrt{\rho}\, \rho' \sqrt{\rho} \} \right )^2$.  At $\lambda_\textsc{b}=4$, the distance between the steady state and the pointer limit  nearly vanishes.

Through the continuous measurement, the environments gain information of the system and the system loses information.  Accordingly, the entropy of the system increases.\cite{Zurek2003,Zurek2002}  As the coupling gets stronger, more information is expected to be lost and thus the entropy goes up monotonically.  Figure \ref{fig:entropy-fidelity} confirms the increase of the von Neumann entropy which converges to the pointer limit (\ref{eq:pointer_limit})  at the strong coupling limit.

\begin{figure}
    \centering
    \includegraphics[width=3in]{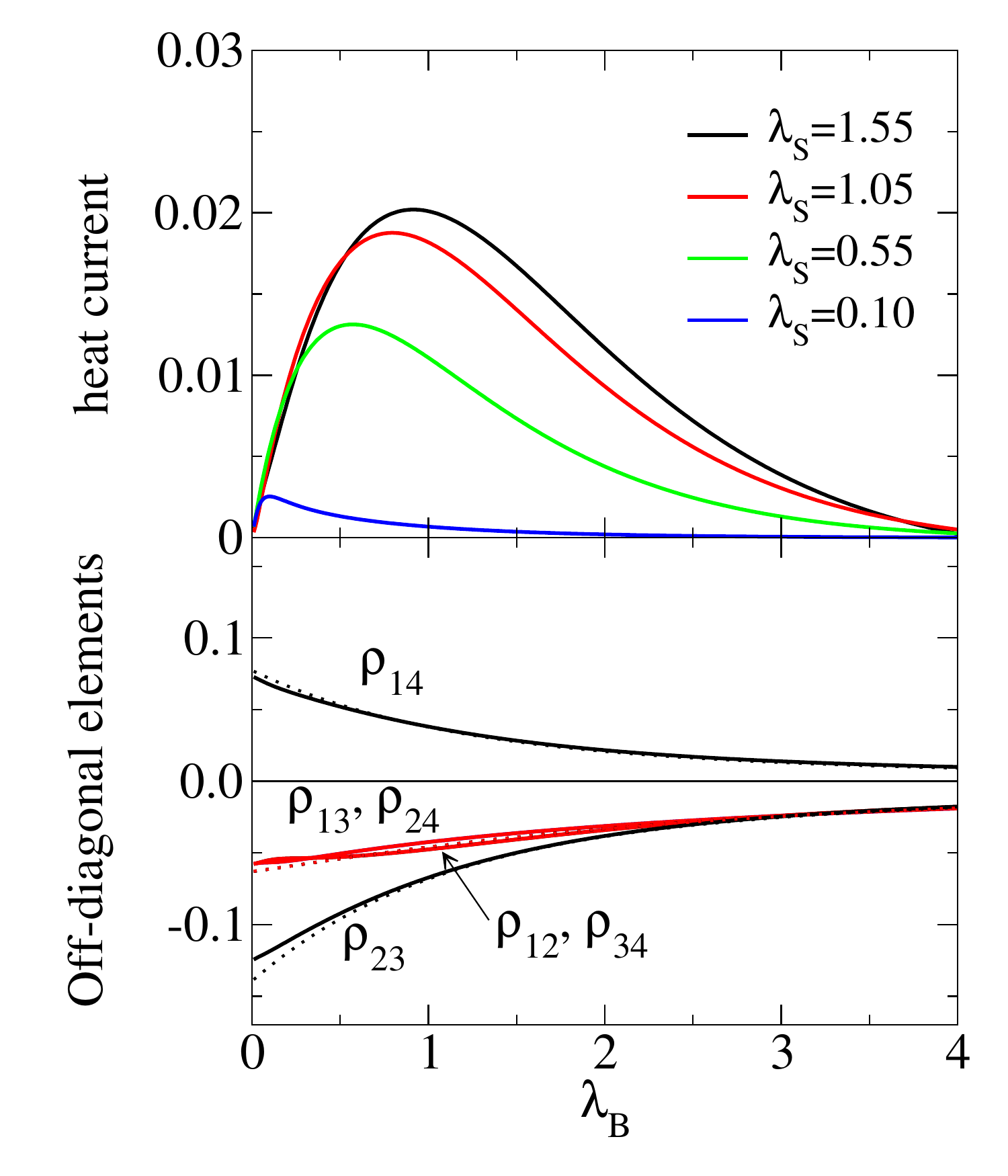}
    \caption{Vanishing heat due to environment-induced decoherence. The upper panel shows the steady-state heat current with $T_{\textsc{b}_1}=2$ and $T_{\textsc{b}_2}=1$.  Notably, the heat current vanishes at the strong coupling limit.  The lower panel shows the decoherence in the pointer basis for $\lambda_\textsc{s}=1.55$.  The dotted lines are the equilibrium density matrix at the effective temperature $T=(T_1+T_2)/2=1.5$.
    The deviation from the equilibrium density is seen only around $\lambda_\textsc{b}=1$ where the heat current reaches its maximum.}
    \label{fig:heat}
\end{figure}

As further evidence of continuous measurement by environments, we also investigated a non-equilibrium steady state.  When different temperatures are used, heat flows through the system.
Heat from the environment B$_\ell$ to the system can be computed as
\begin{equation}\label{eq:heat_def}
J_\ell = -i \Tr_{\textsc{s}} \left \{ \comm{\hat{X}_{\ell}}{\eta_\ell}_{-}\, H_\textsc{s} \right\}.
\end{equation}
Figure \ref{fig:heat} shows the steady state heat current as a function of the coupling strength. In the weak coupling regime, the current increases linearly as expected from the linear response theory.  However, the heat current reaches its maximum and dies off rather quickly as the coupling becomes stronger.  This suppression of heat is predicted earlier as a consequence of the quantum zeno effect\cite{Rebentrost2009} based on a heuristic argument and is observed by Kato-Tanimura\cite{Kato2015}.

The present results show that indeed the decoherence due to environments is responsible for the suppression of heat. The off-diagonal elements of the steady state density look almost identical to those in the stationary state at a single effective temperature $T=(T_1+T_2)/2$  However, there is small but significant difference where the heat current is strong.  The elements $\rho_{13}$ and $\rho_{24}$ deviate from $\rho_{12}$ and $\rho_{34}$ due to the difference in decoherence power between the two environments.  In general a higher temperature environment causes stronger decoherence.\cite{Fleming2012} However, it also depends on the coupling strength as well.  When the coupling strength overcomes  the asymmetry in temperature, the decoherence power of the two environments becomes nearly identical and eventually the asymmetry in the off-diagonal element responsible for the heat conduction vanishes.     

In conclusion, we claim that the ``thermal equilibrium'' of a small quantum system is not the Gibbs state when the coupling to the environments is strong.  Due to continuous measurement by the environment, the stationary state loses the coherency between the pointer states and thus the density is diagonal in the pointer basis rather than in the energy eigenbasis. We further claim that the the stationary state density at the strong coupling limit is the Gibbs state projected onto the pointer basis. The diagonal elements in the pointer basis appear to be insensitive to the coupling strength.  We have demonstrated this proposition by exact numerical calculation using the hierarchical equations of motion.  This strong coupling limit can be used as a bench mark test for analytic models such as the Hamiltonian of mean force. 

\begin{acknowledgments}
We would like to thank Janet Anders, Ala-Nissila, Sahar Alipour, and Erik Aurell for helpful discussion during NORDITA programs.  We also thank James Cresser for interesting discussion.
\end{acknowledgments}

%\bibliographystyle{apsrev4-2}
%\bibliography{references}

\begin{thebibliography}{29}%
   \makeatletter
   \providecommand \@ifxundefined [1]{%
      \@ifx{#1\undefined}
   }%
   \providecommand \@ifnum [1]{%
      \ifnum #1\expandafter \@firstoftwo
      \else \expandafter \@secondoftwo
      \fi
   }%
   \providecommand \@ifx [1]{%
      \ifx #1\expandafter \@firstoftwo
      \else \expandafter \@secondoftwo
      \fi
   }%
   \providecommand \natexlab [1]{#1}%
   \providecommand \enquote  [1]{``#1''}%
   \providecommand \bibnamefont  [1]{#1}%
   \providecommand \bibfnamefont [1]{#1}%
   \providecommand \citenamefont [1]{#1}%
   \providecommand \href@noop [0]{\@secondoftwo}%
   \providecommand \href [0]{\begingroup \@sanitize@url \@href}%
   \providecommand \@href[1]{\@@startlink{#1}\@@href}%
   \providecommand \@@href[1]{\endgroup#1\@@endlink}%
   \providecommand \@sanitize@url [0]{\catcode `\\12\catcode `\$12\catcode
      `\&12\catcode `\#12\catcode `\^12\catcode `\_12\catcode `\%12\relax}%
   \providecommand \@@startlink[1]{}%
   \providecommand \@@endlink[0]{}%
   \providecommand \url  [0]{\begingroup\@sanitize@url \@url }%
   \providecommand \@url [1]{\endgroup\@href {#1}{\urlprefix }}%
   \providecommand \urlprefix  [0]{URL }%
   \providecommand \Eprint [0]{\href }%
   \providecommand \doibase [0]{https://doi.org/}%
   \providecommand \selectlanguage [0]{\@gobble}%
   \providecommand \bibinfo  [0]{\@secondoftwo}%
   \providecommand \bibfield  [0]{\@secondoftwo}%
   \providecommand \translation [1]{[#1]}%
   \providecommand \BibitemOpen [0]{}%
   \providecommand \bibitemStop [0]{}%
   \providecommand \bibitemNoStop [0]{.\EOS\space}%
   \providecommand \EOS [0]{\spacefactor3000\relax}%
   \providecommand \BibitemShut  [1]{\csname bibitem#1\endcsname}%
   \let\auto@bib@innerbib\@empty
   %</preamble>
   \bibitem [{\citenamefont {van Hove}(1957)}]{vanHove1957}%
   \BibitemOpen
   \bibfield  {author} {\bibinfo {author} {\bibfnamefont {L.}~\bibnamefont {van
            Hove}},\ }\href@noop {} {\bibfield  {journal} {\bibinfo  {journal} {Physica
            XXIII}\ } (\bibinfo {year} {1957})}\BibitemShut {NoStop}%
   \bibitem [{\citenamefont {Breuer}\ and\ \citenamefont
      {Petruccione}(2002)}]{Breuer2002}%
   \BibitemOpen
   \bibfield  {author} {\bibinfo {author} {\bibfnamefont {H.-P.}\ \bibnamefont
         {Breuer}}\ and\ \bibinfo {author} {\bibfnamefont {F.}~\bibnamefont
         {Petruccione}},\ }\href@noop {} {\emph {\bibinfo {title} {The Theory of Open
            Quantum Systems}}}\ (\bibinfo  {publisher} {Oxford University Press},\
   \bibinfo {year} {2002})\BibitemShut {NoStop}%
   \bibitem [{\citenamefont {Mori}\ and\ \citenamefont
      {Miyashita}(2008)}]{Mori2008}%
   \BibitemOpen
   \bibfield  {author} {\bibinfo {author} {\bibfnamefont {T.}~\bibnamefont
         {Mori}}\ and\ \bibinfo {author} {\bibfnamefont {S.}~\bibnamefont
         {Miyashita}},\ }\href {https://doi.org/10.1143/JPSJ.77.124005} {\bibfield
      {journal} {\bibinfo  {journal} {Journal of the Physical Society of Japan}\
      }\textbf {\bibinfo {volume} {77}},\ \bibinfo {pages} {124005} (\bibinfo
      {year} {2008})}\BibitemShut {NoStop}%
   \bibitem [{\citenamefont {Genway}\ \emph {et~al.}(2012)\citenamefont {Genway},
      \citenamefont {Ho},\ and\ \citenamefont {Lee}}]{Genway2012}%
   \BibitemOpen
   \bibfield  {author} {\bibinfo {author} {\bibfnamefont {S.}~\bibnamefont
         {Genway}}, \bibinfo {author} {\bibfnamefont {A.~F.}\ \bibnamefont {Ho}},\
      and\ \bibinfo {author} {\bibfnamefont {D.~K.~K.}\ \bibnamefont {Lee}},\
   }\href {https://doi.org/10.1103/PhysRevA.86.023609} {\bibfield  {journal}
      {\bibinfo  {journal} {Phys. Rev. A}\ }\textbf {\bibinfo {volume} {86}},\
      \bibinfo {pages} {023609} (\bibinfo {year} {2012})}\BibitemShut {NoStop}%
   \bibitem [{\citenamefont {Lee}\ \emph {et~al.}(2012)\citenamefont {Lee},
      \citenamefont {Cao},\ and\ \citenamefont {Gong}}]{Lee2012}%
   \BibitemOpen
   \bibfield  {author} {\bibinfo {author} {\bibfnamefont {C.~K.}\ \bibnamefont
         {Lee}}, \bibinfo {author} {\bibfnamefont {J.}~\bibnamefont {Cao}},\ and\
      \bibinfo {author} {\bibfnamefont {J.}~\bibnamefont {Gong}},\ }\href
   {https://doi.org/10.1103/PhysRevE.86.021109} {\bibfield  {journal} {\bibinfo
         {journal} {Phys. Rev. E}\ }\textbf {\bibinfo {volume} {86}},\ \bibinfo
      {pages} {021109} (\bibinfo {year} {2012})}\BibitemShut {NoStop}%
   \bibitem [{\citenamefont {Cai}\ \emph {et~al.}(2014)\citenamefont {Cai},
      \citenamefont {Yang},\ and\ \citenamefont {Sun}}]{Cai2014}%
   \BibitemOpen
   \bibfield  {author} {\bibinfo {author} {\bibfnamefont {C.~Y.}\ \bibnamefont
         {Cai}}, \bibinfo {author} {\bibfnamefont {L.-P.}\ \bibnamefont {Yang}},\ and\
      \bibinfo {author} {\bibfnamefont {C.~P.}\ \bibnamefont {Sun}},\ }\href
   {https://doi.org/10.1103/PhysRevA.89.012128} {\bibfield  {journal} {\bibinfo
         {journal} {Phys. Rev. A}\ }\textbf {\bibinfo {volume} {89}},\ \bibinfo
      {pages} {012128} (\bibinfo {year} {2014})}\BibitemShut {NoStop}%
   \bibitem [{\citenamefont {Xiong}\ \emph {et~al.}(2015)\citenamefont {Xiong},
      \citenamefont {Lo}, \citenamefont {Zhang}, \citenamefont {Feng},\ and\
      \citenamefont {Nori}}]{Xiong2015}%
   \BibitemOpen
   \bibfield  {author} {\bibinfo {author} {\bibfnamefont {H.-N.}\ \bibnamefont
         {Xiong}}, \bibinfo {author} {\bibfnamefont {P.-Y.}\ \bibnamefont {Lo}},
      \bibinfo {author} {\bibfnamefont {W.-M.}\ \bibnamefont {Zhang}}, \bibinfo
      {author} {\bibfnamefont {D.~H.}\ \bibnamefont {Feng}},\ and\ \bibinfo
      {author} {\bibfnamefont {F.}~\bibnamefont {Nori}},\ }\href
   {https://doi.org/10.1038/srep13353} {\bibfield  {journal} {\bibinfo
         {journal} {Scientific Reports}\ }\textbf {\bibinfo {volume} {5}},\ \bibinfo
      {pages} {13353} (\bibinfo {year} {2015})}\BibitemShut {NoStop}%
   \bibitem [{\citenamefont {de~Vega}\ and\ \citenamefont
      {Alonso}(2017)}]{Vega2017}%
   \BibitemOpen
   \bibfield  {author} {\bibinfo {author} {\bibfnamefont {I.}~\bibnamefont
         {de~Vega}}\ and\ \bibinfo {author} {\bibfnamefont {D.}~\bibnamefont
         {Alonso}},\ }\href {https://doi.org/10.1103/RevModPhys.89.015001} {\bibfield
      {journal} {\bibinfo  {journal} {Rev. Mod. Phys.}\ }\textbf {\bibinfo {volume}
         {89}},\ \bibinfo {pages} {015001} (\bibinfo {year} {2017})}\BibitemShut
   {NoStop}%
   \bibitem [{\citenamefont {Gelin}\ and\ \citenamefont
      {Thoss}(2009)}]{Gelin2009}%
   \BibitemOpen
   \bibfield  {author} {\bibinfo {author} {\bibfnamefont {M.~F.}\ \bibnamefont
         {Gelin}}\ and\ \bibinfo {author} {\bibfnamefont {M.}~\bibnamefont {Thoss}},\
   }\href {https://doi.org/10.1103/PhysRevE.79.051121} {\bibfield  {journal}
      {\bibinfo  {journal} {Phys. Rev. E}\ }\textbf {\bibinfo {volume} {79}},\
      \bibinfo {pages} {051121} (\bibinfo {year} {2009})}\BibitemShut {NoStop}%
   \bibitem [{\citenamefont {Campisi}\ \emph {et~al.}(2010)\citenamefont
      {Campisi}, \citenamefont {Zueco},\ and\ \citenamefont
      {Talkner}}]{Campisi2010}%
   \BibitemOpen
   \bibfield  {author} {\bibinfo {author} {\bibfnamefont {M.}~\bibnamefont
         {Campisi}}, \bibinfo {author} {\bibfnamefont {D.}~\bibnamefont {Zueco}},\
      and\ \bibinfo {author} {\bibfnamefont {P.}~\bibnamefont {Talkner}},\ }\href
   {https://doi.org/10.1016/j.chemphys.2010.04.026} {\bibfield  {journal}
      {\bibinfo  {journal} {Chemical Physics}\ }\textbf {\bibinfo {volume} {375}},\
      \bibinfo {pages} {187 } (\bibinfo {year} {2010})}\BibitemShut {NoStop}%
   \bibitem [{\citenamefont {Hilt}\ \emph {et~al.}(2011)\citenamefont {Hilt},
      \citenamefont {Thomas},\ and\ \citenamefont {Lutz}}]{Hilt2011}%
   \BibitemOpen
   \bibfield  {author} {\bibinfo {author} {\bibfnamefont {S.}~\bibnamefont
         {Hilt}}, \bibinfo {author} {\bibfnamefont {B.}~\bibnamefont {Thomas}},\ and\
      \bibinfo {author} {\bibfnamefont {E.}~\bibnamefont {Lutz}},\ }\href
   {https://doi.org/10.1103/PhysRevE.84.031110} {\bibfield  {journal} {\bibinfo
         {journal} {Phys. Rev. E}\ }\textbf {\bibinfo {volume} {84}},\ \bibinfo
      {pages} {031110} (\bibinfo {year} {2011})}\BibitemShut {NoStop}%
   \bibitem [{\citenamefont {Esposito}\ \emph {et~al.}(2015)\citenamefont
      {Esposito}, \citenamefont {Ochoa},\ and\ \citenamefont
      {Galperin}}]{Esposito2015}%
   \BibitemOpen
   \bibfield  {author} {\bibinfo {author} {\bibfnamefont {M.}~\bibnamefont
         {Esposito}}, \bibinfo {author} {\bibfnamefont {M.~A.}\ \bibnamefont
         {Ochoa}},\ and\ \bibinfo {author} {\bibfnamefont {M.}~\bibnamefont
         {Galperin}},\ }\href {https://doi.org/10.1103/PhysRevB.92.235440} {\bibfield
      {journal} {\bibinfo  {journal} {Phys. Rev. B}\ }\textbf {\bibinfo {volume}
         {92}},\ \bibinfo {pages} {235440} (\bibinfo {year} {2015})}\BibitemShut
   {NoStop}%
   \bibitem [{\citenamefont {Seifert}(2016)}]{Seifert2016}%
   \BibitemOpen
   \bibfield  {author} {\bibinfo {author} {\bibfnamefont {U.}~\bibnamefont
         {Seifert}},\ }\href {https://doi.org/10.1103/PhysRevLett.116.020601}
   {\bibfield  {journal} {\bibinfo  {journal} {Phys. Rev. Lett.}\ }\textbf
      {\bibinfo {volume} {116}},\ \bibinfo {pages} {020601} (\bibinfo {year}
      {2016})}\BibitemShut {NoStop}%
   \bibitem [{\citenamefont {Jarzynski}(2017)}]{Jarzynski2017}%
   \BibitemOpen
   \bibfield  {author} {\bibinfo {author} {\bibfnamefont {C.}~\bibnamefont
         {Jarzynski}},\ }\href {https://doi.org/10.1103/PhysRevX.7.011008} {\bibfield
      {journal} {\bibinfo  {journal} {Phys. Rev. X}\ }\textbf {\bibinfo {volume}
         {7}},\ \bibinfo {pages} {011008} (\bibinfo {year} {2017})}\BibitemShut
   {NoStop}%
   \bibitem [{\citenamefont {Miller}\ and\ \citenamefont
      {Anders}(2017)}]{Miller2017}%
   \BibitemOpen
   \bibfield  {author} {\bibinfo {author} {\bibfnamefont {H.~J.~D.}\
         \bibnamefont {Miller}}\ and\ \bibinfo {author} {\bibfnamefont
         {J.}~\bibnamefont {Anders}},\ }\href
   {https://doi.org/10.1103/PhysRevE.95.062123} {\bibfield  {journal} {\bibinfo
         {journal} {Phys. Rev. E}\ }\textbf {\bibinfo {volume} {95}},\ \bibinfo
      {pages} {062123} (\bibinfo {year} {2017})}\BibitemShut {NoStop}%
   \bibitem [{\citenamefont {Strasberg}\ and\ \citenamefont
      {Esposito}(2019)}]{Strasberg2019}%
   \BibitemOpen
   \bibfield  {author} {\bibinfo {author} {\bibfnamefont {P.}~\bibnamefont
         {Strasberg}}\ and\ \bibinfo {author} {\bibfnamefont {M.}~\bibnamefont
         {Esposito}},\ }\href {https://doi.org/10.1103/PhysRevE.99.012120} {\bibfield
      {journal} {\bibinfo  {journal} {Phys. Rev. E}\ }\textbf {\bibinfo {volume}
         {99}},\ \bibinfo {pages} {012120} (\bibinfo {year} {2019})}\BibitemShut
   {NoStop}%
   \bibitem [{\citenamefont {Schlosshauer}(2007)}]{Schlosshauer2007}%
   \BibitemOpen
   \bibfield  {author} {\bibinfo {author} {\bibfnamefont {M.}~\bibnamefont
         {Schlosshauer}},\ }\href@noop {} {\emph {\bibinfo {title} {Decoherence and
            the Quantum-to-Classical Transition}}}\ (\bibinfo  {publisher} {Springer},\
   \bibinfo {year} {2007})\BibitemShut {NoStop}%
   \bibitem [{\citenamefont {Zurek}(2003)}]{Zurek2003}%
   \BibitemOpen
   \bibfield  {author} {\bibinfo {author} {\bibfnamefont {W.~H.}\ \bibnamefont
         {Zurek}},\ }\href {https://doi.org/10.1103/RevModPhys.75.715} {\bibfield
      {journal} {\bibinfo  {journal} {Rev. Mod. Phys.}\ }\textbf {\bibinfo {volume}
         {75}},\ \bibinfo {pages} {715} (\bibinfo {year} {2003})}\BibitemShut
   {NoStop}%
   \bibitem [{\citenamefont {Eisert}(2004)}]{Eisert2004}%
   \BibitemOpen
   \bibfield  {author} {\bibinfo {author} {\bibfnamefont {J.}~\bibnamefont
         {Eisert}},\ }\href {https://doi.org/10.1103/PhysRevLett.92.210401} {\bibfield
      {journal} {\bibinfo  {journal} {Phys. Rev. Lett.}\ }\textbf {\bibinfo
         {volume} {92}},\ \bibinfo {pages} {210401} (\bibinfo {year}
      {2004})}\BibitemShut {NoStop}%
   \bibitem [{\citenamefont {Jack}\ and\ \citenamefont
      {Collett}(2000)}]{Jack2000}%
   \BibitemOpen
   \bibfield  {author} {\bibinfo {author} {\bibfnamefont {M.~W.}\ \bibnamefont
         {Jack}}\ and\ \bibinfo {author} {\bibfnamefont {M.~J.}\ \bibnamefont
         {Collett}},\ }\href {https://doi.org/10.1103/PhysRevA.61.062106} {\bibfield
      {journal} {\bibinfo  {journal} {Phys. Rev. A}\ }\textbf {\bibinfo {volume}
         {61}},\ \bibinfo {pages} {062106} (\bibinfo {year} {2000})}\BibitemShut
   {NoStop}%
   \bibitem [{\citenamefont {Ashida}\ \emph {et~al.}(2018)\citenamefont {Ashida},
      \citenamefont {Saito},\ and\ \citenamefont {Ueda}}]{Ashida2018}%
   \BibitemOpen
   \bibfield  {author} {\bibinfo {author} {\bibfnamefont {Y.}~\bibnamefont
         {Ashida}}, \bibinfo {author} {\bibfnamefont {K.}~\bibnamefont {Saito}},\ and\
      \bibinfo {author} {\bibfnamefont {M.}~\bibnamefont {Ueda}},\ }\href
   {https://doi.org/10.1103/PhysRevLett.121.170402} {\bibfield  {journal}
      {\bibinfo  {journal} {Phys. Rev. Lett.}\ }\textbf {\bibinfo {volume} {121}},\
      \bibinfo {pages} {170402} (\bibinfo {year} {2018})}\BibitemShut {NoStop}%
   \bibitem [{Note1()}]{Note1}%
   \BibitemOpen
   \bibinfo {note} {If two qubits share the same environment, decoherence-free
      subspaces could be formed, which is protected from decoherence due to
      symmetry. We avoid the decoherence free subspace by using two independent
      environments.}\BibitemShut {Stop}%
   \bibitem [{\citenamefont {Ishizaki}\ and\ \citenamefont
      {Fleming}(2009)}]{Ishizaki2009}%
   \BibitemOpen
   \bibfield  {author} {\bibinfo {author} {\bibfnamefont {A.}~\bibnamefont
         {Ishizaki}}\ and\ \bibinfo {author} {\bibfnamefont {G.~R.}\ \bibnamefont
         {Fleming}},\ }\href@noop {} {\bibfield  {journal} {\bibinfo  {journal} {J.
            Chem. Phys.}\ }\textbf {\bibinfo {volume} {130}},\ \bibinfo {eid} {234111}
      (\bibinfo {year} {2009})}\BibitemShut {NoStop}%
   \bibitem [{\citenamefont {Kato}\ and\ \citenamefont
      {Tanimura}(2015)}]{Kato2015}%
   \BibitemOpen
   \bibfield  {author} {\bibinfo {author} {\bibfnamefont {A.}~\bibnamefont
         {Kato}}\ and\ \bibinfo {author} {\bibfnamefont {Y.}~\bibnamefont
         {Tanimura}},\ }\href {https://doi.org/10.1063/1.4928192} {\bibfield
      {journal} {\bibinfo  {journal} {The Journal of Chemical Physics}\ }\textbf
      {\bibinfo {volume} {143}},\ \bibinfo {pages} {064107} (\bibinfo {year}
      {2015})}\BibitemShut {NoStop}%
   \bibitem [{\citenamefont {Kato}\ and\ \citenamefont
      {Tanimura}(2016)}]{Kato2016}%
   \BibitemOpen
   \bibfield  {author} {\bibinfo {author} {\bibfnamefont {A.}~\bibnamefont
         {Kato}}\ and\ \bibinfo {author} {\bibfnamefont {Y.}~\bibnamefont
         {Tanimura}},\ }\href {https://doi.org/10.1063/1.4971370} {\bibfield
      {journal} {\bibinfo  {journal} {The Journal of Chemical Physics}\ }\textbf
      {\bibinfo {volume} {145}},\ \bibinfo {pages} {224105} (\bibinfo {year}
      {2016})}\BibitemShut {NoStop}%
   \bibitem [{\citenamefont {Xu}\ \emph {et~al.}(2009)\citenamefont {Xu},
      \citenamefont {Tian}, \citenamefont {Xu}, \citenamefont {Shi},\ and\
      \citenamefont {Yan}}]{Xu2009}%
   \BibitemOpen
   \bibfield  {author} {\bibinfo {author} {\bibfnamefont {R.-X.}\ \bibnamefont
         {Xu}}, \bibinfo {author} {\bibfnamefont {B.-L.}\ \bibnamefont {Tian}},
      \bibinfo {author} {\bibfnamefont {J.}~\bibnamefont {Xu}}, \bibinfo {author}
      {\bibfnamefont {Q.}~\bibnamefont {Shi}},\ and\ \bibinfo {author}
      {\bibfnamefont {Y.}~\bibnamefont {Yan}},\ }\href
   {https://doi.org/http://dx.doi.org/10.1063/1.3268922} {\bibfield  {journal}
      {\bibinfo  {journal} {The Journal of Chemical Physics}\ }\textbf {\bibinfo
         {volume} {131}},\ \bibinfo {eid} {214111} (\bibinfo {year}
      {2009})}\BibitemShut {NoStop}%
   \bibitem [{\citenamefont {Zurek}(2002)}]{Zurek2002}%
   \BibitemOpen
   \bibfield  {author} {\bibinfo {author} {\bibfnamefont {W.~H.}\ \bibnamefont
         {Zurek}},\ }\href@noop {} {\emph {\bibinfo {title} {Decoherence and the
            Transition from Quantum to Classical -- Revisited}}},\ \bibinfo {type} {Tech.
      Rep.}\ (\bibinfo  {institution} {Los Alamos National Laboratory},\ \bibinfo
   {year} {2002})\BibitemShut {NoStop}%
   \bibitem [{\citenamefont {Rebentrost}\ \emph {et~al.}(2009)\citenamefont
      {Rebentrost}, \citenamefont {Mohseni}, \citenamefont {Kassal}, \citenamefont
      {Lloyd},\ and\ \citenamefont {Aspuru-Guzik}}]{Rebentrost2009}%
   \BibitemOpen
   \bibfield  {author} {\bibinfo {author} {\bibfnamefont {P.}~\bibnamefont
         {Rebentrost}}, \bibinfo {author} {\bibfnamefont {M.}~\bibnamefont {Mohseni}},
      \bibinfo {author} {\bibfnamefont {I.}~\bibnamefont {Kassal}}, \bibinfo
      {author} {\bibfnamefont {S.}~\bibnamefont {Lloyd}},\ and\ \bibinfo {author}
      {\bibfnamefont {A.}~\bibnamefont {Aspuru-Guzik}},\ }\href
   {https://doi.org/10.1088/1367-2630/11/3/033003} {\bibfield  {journal}
      {\bibinfo  {journal} {New Journal of Physics}\ }\textbf {\bibinfo {volume}
         {11}},\ \bibinfo {pages} {033003} (\bibinfo {year} {2009})}\BibitemShut
   {NoStop}%
   \bibitem [{\citenamefont {Fleming}\ \emph {et~al.}(2012)\citenamefont
      {Fleming}, \citenamefont {Hu},\ and\ \citenamefont {Roura}}]{Fleming2012}%
   \BibitemOpen
   \bibfield  {author} {\bibinfo {author} {\bibfnamefont {C.}~\bibnamefont
         {Fleming}}, \bibinfo {author} {\bibfnamefont {B.}~\bibnamefont {Hu}},\ and\
      \bibinfo {author} {\bibfnamefont {A.}~\bibnamefont {Roura}},\ }\href
   {https://doi.org/10.1016/j.physa.2012.04.008} {\bibfield  {journal} {\bibinfo
         {journal} {Physica A: Statistical Mechanics and its Applications}\ }\textbf
      {\bibinfo {volume} {391}},\ \bibinfo {pages} {4206 } (\bibinfo {year}
      {2012})}\BibitemShut {NoStop}%
\end{thebibliography}
%apsrev4-2.bst 2019-01-14 (MD) hand-edited version of apsrev4-1.bst
%Control: key (0)
%Control: author (72) initials jnrlst
%Control: editor formatted (1) identically to author
%Control: production of article title (-1) disabled
%Control: page (0) single
%Control: year (1) truncated
%Control: production of eprint (0) enabled
%

\end{document}